\documentclass[12pt]{article}
\usepackage{graphicx}
\newcommand{\beq}{\begin{equation}}
\newcommand{\eeq}{\end{equation}}
\newcommand{\bea}{\begin{eqnarray}}
\newcommand{\eea}{\end{eqnarray}}

\def\fun#1#2{\lower3.6pt\vbox{\baselineskip0pt\lineskip.9pt
  \ialign{$\mathsurround=0pt#1\hfil##\hfil$\crcr#2\crcr\sim\crcr}}}

\begin{document}
\begin{titlepage}
\begin{flushleft}
       \hfill                      {\tt hep-th/0305150}\\
       \hfill                       FIT HE - 03-03 \\
\end{flushleft}
\vspace*{3mm}
\begin{center}
{\bf\LARGE Instability of thick brane worlds \\ }
\vspace*{5mm}

\bigskip

{\large Kazuo Ghoroku\footnote{\tt gouroku@dontaku.fit.ac.jp}\\ }
\vspace*{1mm}
{
\large 
Fukuoka Institute of Technology, Wajiro, Higashi-ku}\\
{
\large 
Fukuoka 811-0295, Japan\\}
\vspace*{3mm}

%{\large Akihiro Nakamura\footnote{\tt nakamura@sci.kagoshima-u.ac.jp}}\\
%\vspace*{1mm}
%{\large Department of Physics, Kagoshima University, Korimoto 1-21-35}\\
%{\large Kagoshima 890-0065, Japan\\}
%\vspace*{3mm}

{\large Masanobu Yahiro\footnote{\tt yahiro@sci.u-ryukyu.ac.jp} \\}
\vspace{1mm}
{
\large 
Department of Physics and Earth Sciences, University of the Ryukyus,
Nishihara-chou, Okinawa 903-0213, Japan \\}

\vspace*{10mm}

\end{center}

\begin{abstract}
 We examined 5d thick brane worlds constructed by a real scalar field.
The solutions are obtained in terms of a simple form of smooth warp factor.
For the case of dS brane, we found a disease of the solutions. For example,
it is impossible to construct thick-brane worlds with our simple 
smoothing. This result is independent of supersymmetry or other symmetries.

\end{abstract}
\end{titlepage}

\section{Introduction}

After the proposal of a thin brane world given in \cite{RS1}, the 
construction and the use of smooth brane world solutions in higher dimensions
have been given by many authors \cite{DFGK,CEHS}. This direction seems to be
natural from the viewpoint of field theory since the thin brane should be 
considered as a limit of a thick brane
which would be obtained as a soliton solution of a higher dimensional
field theory coupled with gravity. An interesting and simple case 
in obtaining a thick brane solution is the
5d gravity with a single real scalar. For supergravity, it is known as
a difficult task to find such soliton solutions \cite{KL,BC2,KL2,GL,DLS}, but
many kinds of thick brane solutions are possible
in non-supersymmetric case \cite{DFGK,CEHS,M,KT,IS,DF,Gio,KKSoda}. 
And various problems, particularly 
the localization of gravity, have been studied in terms of the latter model. 

On the other hand, the fluctuation of the
scalar field which constructs the thick brane has not been studied
enough. In \cite{DFGK}, it was pointed out that,
in the thin brane limit, the bulk mass of this field becomes very
large and the fluctuation would be frozen leaving only the thin brane as a 
background configuration. However before taking this
limit, we should carefully study the spectrum of this fluctuation.
The classical solution for this scalar has generally
a form of kink along the
fifth coordinate, and the second derivative of its potential is
negative around the kink or near the brane. This property would be
common to all thick brane models. 

    In this paper, we consider two types of thick branes, 
dS and or Poincar{\'e} branes. 
For the thick dS brane case, we show that it is not 
possible to constuct such a brane with a real scalar field. 
For the thick Poincar{\'e} brane case, we show that the scalar fluctuation has 
a tachyon, when mixing with the scalar fluctuation of the metric is 
ignored. This tachyonic mode might be removed
when the mixing is taken into account \cite{DF,Gio,KKSoda}. 

\vspace{.2cm}
In Section 2, we set our model with a real scalar and show thick brane 
solutions. In Section 3, the stability of the thick brane worlds
is examined. The final section is devoted to conclusion.

\section{Thick brane solution}

We begin with the action for five-dimensional gravity coupled 
with a real scalar ($\phi$),
\bea
   S_g=\int d^4\!xdy\sqrt{-g}
   \left\{{1\over 2\kappa^2}R 
    -{1\over 2}(\partial\phi)^2-V(\phi)\right\}   \ ,
                                                     \label{acg}
\eea
where $\kappa^2$ denotes the five-dimensional gravitational constant,
and the form of potential $V$ is not specified at this stage.
The equations of motion derived from the action (\ref{acg}) are 
\bea
  &&  G_{MN}=
      \kappa^2\left\{
    \partial_M\phi\partial_N\phi-g_{MN}
    \left({1\over 2}(\partial\phi)^2+V\right)\right\} \ , \label{eins}
\\
  &&\qquad\qquad
      {1\over\sqrt{-g}}\partial_M\left\{\sqrt{-g}g^{MN}\partial_N\phi\right\}
      ={\partial V\over \partial\phi} \ .   \label{dila}
\eea
We assume metric of the form 
\bea
   ds^2=g_{MN}dx^M dx^N=A^2(y)(-dt^2+a_0^2(t)\gamma_{ij}dx^i dx^j)
         +dy^2 \ ,                                 \label{fmet}
\eea
where $\gamma_{ij}=(1+k\delta_{mn}x^m x^n/4)^{-2}\delta_{ij}$.
The 3d scale factor $a_0(t)$ is soluble for each $k$, for example 
$a_0=e^{\sqrt{\lambda} t}$ in the case of $k=0$. 
We also assume that $\phi$ depends only on the fifth coordinate $y$. 
These assumptions simplify the equations of motion, 
(\ref{eins}) and (\ref{dila}), as 
\bea
   {A''\over A}+\left({A'\over A}\right)^2
     -{\lambda \over A^2}
  &=&-{\kappa^2\over 3}
       \left({1\over 2}(\phi')^2 +V(\phi)\right) \ ,             
\label{tteq}
\\
          \left({A'\over A}\right)^2-{\lambda \over A^2} 
    &=&{\kappa^2\over 6}
     \left({1\over 2}(\phi')^2 -V(\phi)\right) \ ,
                                                  \label{yyeq}
\\
              \phi''+4\phi'{A'\over A}
        &=&{\partial V\over \partial \phi} \ ,  \label{seq}
\eea
where $'=d/dy$. From Eqs.(\ref{tteq}) and (\ref{yyeq}), we obtain
\beq
 (\phi')^2=-{3\over \kappa^2}\left({A''\over A}-({A'\over A})^2
                   +{\lambda\over A^2}\right) \ , \label{secderi}
\eeq
\beq
 V=-{9\over 2\kappa^2}\left({A''\over 3A}+({A'\over A})^2
                   -{\lambda\over A^2}\right) \ . \label{potential}
\eeq
The $\phi'$ and $V$ defined by Eqs. (\ref{secderi}) and (\ref{potential}) 
automatically satisfy Eq. (\ref{seq}), so Eq. (\ref{seq}) is redundant.
If $A$ is given and smooth everywhere, 
$\phi'$ and $V$ are obtained from the $A$ and its derivative. 
The scalar field thus obtained forms a thick brane world.

\vspace{.5cm}
The warp factor  of the thin brane world is written 
as $A=A(|y|)$ \cite{bre} and singular at $y=0$, the position of the
thin brane. Smoothing out the singularity corresponds to introducing 
the thickness to the brane. Here we introduce a smooth warp factor of 
the form $A=A(f(b,y))$ with 
\beq
  f(b,y)={1\over b}\log (\cosh (b y)) \ , 
\label{f}
\eeq
replacing $|y|$ by $f$ without changing the functional form of $A(|y|)$. 
This replacement was 
adopted also in \cite{CEHS}, and the original thin brane is recovered 
in the large limit of parameter $b$, since $f\to |y|$ as $b\to \infty$.
Although there are many other smoothed forms and parametrizations
\cite{DFGK,M,KT,IS,KKSoda},
we take (\ref{f}) for convenience.

\vspace{.5cm}
\section{Instability of thick dS brane}

The stability of the thick brane solutions is investigated in two manners. 
First we check the reality of $\phi$ through (\ref{secderi}).

\vspace{.5cm}
For $\lambda>0$ and $\Lambda<0$, we have a warp factor $A$ of the form 
\beq
     A(y)={\sqrt{\lambda}\over \mu}\sinh\left[\mu(y_H-f(y))\right] , 
\eeq
leading to 
\beq
  (\phi')^2={3\mu^2 \over \kappa^2}{{b\over 2\mu}\sinh[2\mu(y_H-f)]-1\over 
             \cosh^2(by)\sinh^2[\mu(y_H-f)]} \ ,
\eeq
where $\mu=\sqrt{-\Lambda/6}$ and 
$\sinh\left[\mu y_H \right]={\mu/\sqrt{\lambda}}$. 
Obviously, there exists a region of negative $(\phi')^2$ for $y$ 
satisfying $\sinh[2\mu(y_H-f)]<2\mu/b$, near the extended horizon
$y=\hat{y}_H >y_H$ defined by $A(f(\hat{y}_H))=0$. 
Thus, the thick-brane solution is unstable in the region. 
A similar analysis is applicable for the case of $\lambda>0$ and $\Lambda>0$, 
and the same result is obtained also for this case. 
The unstable region disappears in the limit 
of either $b\to \infty$ or $\lambda\to 0$. 
The former corresponds to the thin-brane limit, and 
the latter discussed below does to the limit of $y_H \to \infty$.

\vspace{.5cm}
As for ${\lambda}=0$ and ${\Lambda}<0$, we have
$$A(y)=e^{-\mu f(y)}$$ 
\beq
(\phi')^2={3\mu \over \kappa^2}{b\over \cosh^2(by)} >0 \ .
\label{phi-M4}
\eeq
Thus, the reality condition for $\phi$ is satisfied. 
Solving (\ref{phi-M4}), we get 
\beq
   \phi=v \arctan{[\tanh{{by \over 2}}]} , \quad
   v= \sqrt{{12 \mu \over \kappa^2 b}} \ , 
\eeq
and eventually reach to the so-called sine-Gordon potential 
\beq
   V={b^2 v^2 \over 16}\{1+\cos{{4\phi \over v}}- 
{\kappa^2 v^2\over 3}(1-\cos{{4\phi \over v}}) 
  \}  \ . \label{singoldon}
\eeq
Note that the potential minimum at $\phi=\pm \pi v /4$ is negative 
for finite $\kappa^2$. This implies that the AdS$_5$ bulk configuration 
is recovered in the thin brane limit. 
Further, we notice that the part proportional to
$\kappa^2$ is negative for any $\phi$ and it provides an 
attractive force for the fluctuation of $\phi$ 
from its classical configuration, the kink.

\vspace{.5cm}
Now, we consider the second point to be examined. 
In general, the stability of a field configuration
is assured by the positivity of the second derivative of the potential
around the configuration. For the case of $\lambda=0$, it turns out to be 
\beq
   {d^2V\over d\phi^2}=b^2(1+4{\mu\over b}){\sinh^2(by)-1\over \cosh^2(by)}\equiv M^2(y) 
\ ,
\eeq
indicating that $M^2(y)$ is negative near the center of the thick 
brane, $y\sim 0$. It is a non-trivial problem to see whether this region
causes the instability of the thick brane solution or not. 
On the other hand, we should notice another important point.
The scalar fluctuation mixes in general 
with the scalar components of the metric \cite{DF,Gio, KKSoda}, 
when the scalar takes a nontrivial background configuration
just as the case considered here. 

\vspace{.5cm}
In order to show the importance of the mixing, we first switch it off.
The resultant eigenvalue equation for the fluctuation
($\delta \phi$)
of $\phi$ around its classical solution is obtained,  by representing it 
as $\delta \phi=\int dm \varphi (m,x)\Phi_m(y)$, as  
\beq
  \Phi_m''+{4A'\over A}\Phi_m'+{m^2\over A^2}\Phi_m=M^2\Phi_m \ .
    \label{fluctu}
\eeq
Before performing a numerical analysis, we make a small speculation
on this equation. By introducing the variable $q=by$ and the parameter
$\xi=\mu/b=v^2\kappa^2/12$, (\ref{fluctu}) is written as
\beq
  \ddot{\Phi}_m+{4\dot{A}\over A}\dot{\Phi}_m+{m^2\over b^2A^2}\Phi_m=
\tilde{M}^2\Phi_m \ ,
    \label{fluctu2}
\eeq
where
$$ A=\cosh^{-\xi}(q),\qquad
\tilde{M}^2=(1+4{\xi}){\sinh^2(q)-1\over \cosh^2(q)}. $$
where the dot $\dot{}$ denotes $\partial_q$.
Equation (\ref{fluctu2}) shows that its eigenvalue $m^2/b^2$ depends only on 
$\xi$: $m^2/b^2=P(\xi)$. In the limit of small $\kappa^2$ 
with $v$ and $b$ fixed, $A$ tends to 1. 
In such a flat space-time, as a well-known fact, 
$\delta \phi$ has a zero mode
leading to $P(0)=0$ at $\xi=0$.  
At small $\xi$,  $m^2$ is then represented as 
\beq
          m^2= c \xi b^2= c \mu b 
\eeq
with a constant $c$. Figure 1 shows that numerical solution 
yields $c=-2.73$. 
The fact of $c<0$ is understandable, as shown below, through the 
the Shroedinger-type eigenvalue equation derived from (\ref{fluctu2}), 
since in the eigenvalue equation the potential becomes more attractive 
near the brane when $\xi$ is put on. 

\vspace{.5cm}
Here we take the boundary condition for $\Phi_m$ as $\Phi_m'(0)=0$,
which is corresponding to the Neuman condition used for the Planck brane.
While $\Phi_m(0)$ is arbitrary due to the linearity of the equation.
The ground state wave function is shown in the Fig.2. It is localized on the
brane.

%%%%%%%%%%%%%%% Fig %%%%%%%%%%%%%%
\begin{figure}[htbp]
\begin{center}
\voffset=15cm
\includegraphics[width=9cm,height=7cm]{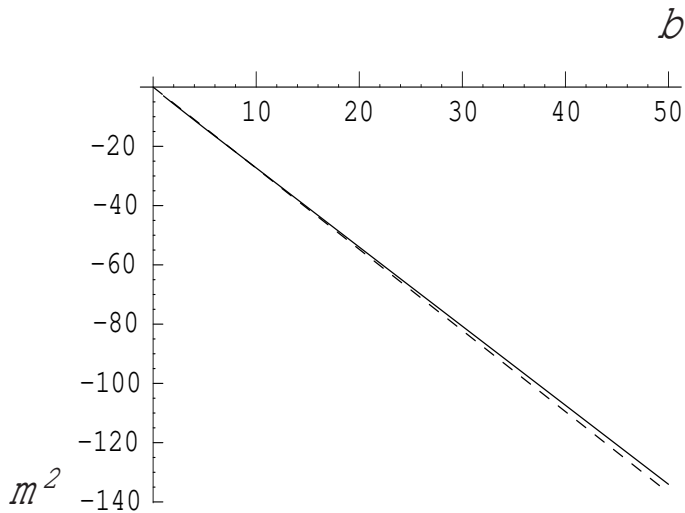} 
\caption{Tachyonic eigenemass as a function of $b$. 
Here $\mu$ is fixed at 1. 
The solid line shows the exact solution calculated numerically, 
while the dashed one does a line $m^2=-2.73 b$.
\label{mxigraph}}
\end{center}
\end{figure}
%%%%%%%%%%%%%%% Fig %%%%%%%%%%%%%%

\vspace{.5cm}
The results shown above could be understood more qualitatively 
according to the usual formulation. 
Rewrite (\ref{fluctu}), with new quantities $z$ and $u(z)$  
defined by $\partial z/\partial y=\pm A^{-1}$
and $\Phi_m=A^{-3/2}u(z)$, as 
\beq
 [-\partial_z^2+V_g(z)]u(z)=m^2 u(z) , \ \label{warp3}
\eeq
where the potential $V_g(z)$ is given as
\beq
 V_g(z)={9\over 4}(A')^2+{3\over 2}AA''+A^2M^2.
\eeq 
The first two terms of the potential $V_g(z)$ are the pure gravitational
part and they would be important in obtaining 
the tachyonic state given above. Further the finiteness of $\kappa$
in the third term is also important. 
To see the total gravitational effects, we compare
the potential $V_g$ with the case of $\kappa=0$ (denoted by $V_0$).
In Fig.2, we show them for $\mu=b=1$. For the case of $V_0$, the lowest
eigenvalue is seen at zero. While the potential becomes deep and its shape
is the famous volcano type when the gravitational effect works. 
As a result, the ground 
state energy is pull down to the tachyonic point 
which is shown in the Fig.2 by $E_g$. These
results are naturally understood.

\vspace{.3cm}
Now we consider the mixing of the scalar fluctuation with the metric ones. 
In this case, the potential $V_g$ in
the corresponding Schr{\"o}dinger-like equation (\ref{warp3}) is modified
and it changes to a positive-definite 
one \cite{DF,Gio,KKSoda}, so that the scalar fluctuation has no tachyon.

%%%%%%%%%%%%%%% Fig %%%%%%%%%%%%%%
\begin{figure}[htbp]
\begin{center}
\voffset=15cm
\includegraphics[width=9cm,height=7cm]{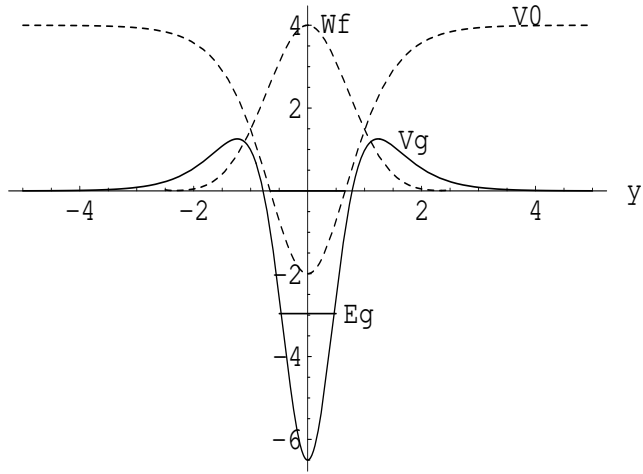} 
\caption{Potential $V_g$ are shown for $\mu=b=1$. The case of
$\kappa=0$ is shown by the dotted curve denoted by $V_0$, and
the point denoted by $E_g=-2.963$ is the lowest eigenvalue with the
potential $V_g$. By the dotted curve Wf, the ground state wave
function is expressed.
\label{mxigraph}}
\end{center}
\end{figure}
%%%%%%%%%%%%%%% Fig %%%%%%%%%%%%%%

\section{Conclusion}

   We have studied, through two conditions, whether 
the thick-brane world constructed from the real scalar field  is 
stable or not.  For the thick dS brane case, 
with a positive 4d cosmological constant ($\lambda>0$), 
the thick-brane world breaks, 
near the horizon, the reality condition of the scalar field 
($\phi$). 
%%%%%%% KG
This difficulty is cured by taking the limit of $\lambda\to 0$ or
the thin brane limit. When we consider the Poincar{\'e} brane 
with no $\lambda$, on the other hand, the thick-brane world recovers 
the reality condition, and the fluctuation of $\phi$ does not generate 
any tachyonic mode by an effect of the coupling between the scalar and metric 
fluctuations. Hence, we can conclude that only 
the thick dS brane is unstable. 

\vspace{.5cm}
    The configuration of the thick brane constructed above, in general, 
depends on how to smooth the warp factor $A(y)$ of the thin-brane world. 
A different type of smoothing is proposed in \cite{KKSoda}. 
In this type of smoothing, 
the stability of thick dS brane may depend on parameters taken; 
the thick brane seems to be unstable for large thickness of the brane, 
that is, for $n<1$ in the notation of \cite{KKSoda}. 
Further analysis is thus highly expected.

\section*{Acknowledgments}
This work has been supported in part by the Grants-in-Aid for
Scientific Research (13135223, 14540271)
of the Ministry of Education, Science, Sports, and Culture of Japan.


\vspace{.5cm}
\begin{thebibliography}{99}
\bibitem{RS1} L. Randall and R. Sundrum, Phys. Rev. Lett. {\bf 83} (1999)
3370 ({\tt hep-ph/9905221}): Phys. Rev. Lett. {\bf 83} (1999)
4690, ({\tt hep-th/9906064}).

\bibitem{DFGK} O. DeWolfe, D.Z. Freedman, S.S. Gubser and A. Karch.
 Phys. Rev. {\bf D62} 046008, 2000, ({\tt hep-th/9909134}). 
\bibitem{CEHS} C. Csaki, J. Erlich, T.J. Hollowood and Y. Shirman, 
Nucl. Phys. {\bf B581} 309-338, 2000
({\tt hep-th/0001033}).

%No go theorem in super gravity

\bibitem{KL} R.~Kallosh and A.~Linde, JHEP {\bf 0002} (2000) 005
  ({\tt hep-th/0001071}).
\bibitem{BC2} K.~Behrndt and M.~Cvetic, Phys. Rev. {\bf D61} (2000)
  101901, ({\tt hep-th/0001159}).
\bibitem{KL2} R.~Kallosh and A.~Linde, ({\tt hep-th/9910021}).
\bibitem{GL} G.W.~Gibbons and N.D.~Lambert,  Phys. Lett, {\bf B488}
(2000) 90, ({\tt hep-th/0003197}).
\bibitem{DLS} M.J. Duff, J.T. Liu and and K.S.~Stelle, 
({\tt hep-th/0007120}).
%%%%%%%%%%%%%%%%%%%%%%%%%%%%%%%%%%


\bibitem{M} M. Gremm, Phys. Lett. {\bf B478} 434, 2000
({\tt hep-th/9912060}):  Phys.Rev. {\bf D62} 044017, 2000
({\tt hep-th/0002040})
\bibitem{KT} A. Kehagias and K. Tamvakis, Mod. Phys. Lett. {\bf A17} 1767,
2002, ({\tt hep-th/0011006}) 

%2)  SOME PROPERTIES OF DOMAIN WALL SOLUTION IN THE RANDALL-SUNDRUM MODEL.
\bibitem{IS} S. Ichinose, Class. Quant. Grav. {\bf 18} 5239, 2001
({\tt hep-th/0107254}) 

% mixing term 
\bibitem{DF} O. DeWolfe and D.Z. Freedman, {\tt hep-th/0002226}. 

\bibitem{Gio} M. Giovannini, Phys. Rev. {\bf D64} (2001) 064023. 

\bibitem{KKSoda} S. Kobayashi, K. Koyama and J. Soda, 
	Phys. Rev. {\bf D65} (2002) 064014. 


%%%%%%%%%%%%%%%%%%% New %%%%%%%%%%%%%

%**************************************************************
\bibitem{bre}
I. Brevik, K. Ghoroku, S. D. Odintsov and M. Yahiro, Phys. Rev. {\bf 66}
   (2002) 064016, ({\tt hep-th/0204066}). 


%%%%%%%%%%%%%%%%%%%%%%%%%%%%%%%%%%%%%%%%%%%%%%%%%%%%%%%%%%%%%%%%%%%%%%%%%%



\end{thebibliography}
\end{document}